\documentclass[11pt,letterpaper]{article}
\pdfoutput=1
\usepackage{amscd}
\usepackage{amsmath, bbm} 

\usepackage{jheppub} 
\usepackage{slashed}
\usepackage{multirow}
\usepackage{array}
\usepackage{graphicx}
\usepackage{tikz-feynman}
\usepackage{tikz-cd}
\usepackage{longtable}
\usepackage{setspace}
\allowdisplaybreaks
\newcommand\mybox[1]{\parbox[t]{0.9\textwidth}{%
    \setstretch{1.33}\raggedright$\displaystyle #1$}}

\title{Lorentz and permutation invariants of particles I} 

\author[a]{Ben Gripaios,}
\author[b]{Ward Haddadin,}
\author[a]{and Christopher G. Lester}
\affiliation[a]{Cavendish Laboratory, University of Cambridge, J.J. Thomson Avenue, Cambridge, CB3 0HE, United Kingdom} 
\affiliation[b]{DAMTP, University of Cambridge, Wilberforce Road, Cambridge, CB3 0WA, United Kingdom}

\emailAdd{gripaios@hep.phy.cam.ac.uk}
\emailAdd{w.haddadin@damtp.cam.ac.uk}
\abstract{A theorem of Weyl tells us that the Lorentz (and parity)
  invariant polynomials in the momenta of $n$ particles are generated
  by the dot products. We extend this result to include the action of
  an arbitrary permutation group $P \subset S_n$ on the particles,
 to take account of the quantum-field-theoretic fact that particles
 can be indistinguishable. Doing so provides a convenient set of
 variables for describing scattering processes involving identical
 particles, such as $pp \to jjj$, for which we provide an explicit set
 of Lorentz and permutation invariant generators.}

\newcommand{\C}{\mathbb{C}}
\newcommand{\Z}{\mathbb{Z}}

\newcommand{\im}{\mathrm{im}\,}
\begin{document} 
\maketitle
\flushbottom

\section{Introduction}\label{sec:Intro}
Given the momentum vectors $p_i$ of $n$ particles in $d$ spacetime dimensions, an old theorem of Weyl \cite{Weyl} tells us that the Lorentz- and parity-invariant polynomials are generated by the dot products $p_i \cdot p_j$.\footnote{If parity is not a symmetry, we must also include the polynomials
obtained by contracting $d$ momenta with the Levi-Civita tensor, a
complication whose discussion we postpone until \S \ref{sec:Parity}.} 

This theorem (or rather its obvious generalization from polynomials to
the field of rational functions, the ring of formal power series, and
thence to the whole gamut of functions typically considered in
physics) has become so ubiquitous that it is, by and large, taken for
granted nowadays. But it is  perhaps in need of a makeover, given what we know about quantum field theory, namely that the particles that correspond to excitations of a single quantum field are indistinguishable. 

Thus, supposing that some subsets of our $n$ particles are identical
(e.g. in a process in which two protons at the LHC collide to produce
three jets), it is apposite to consider not just
arbitrary Lorentz-invariant polynomials, but rather to restrict to those that are, in addition, invariant under the group of
permutations of the identical particles (e.g. $S_2 \times S_3$ in our $pp
\to jjj$ example).

To give a first explicit example of why this might be helpful in phenomenological
analyses, it is useful to consider the situation in which the analysis
is carried out, as is increasingly the case, by a supremely {\em
  un}intelligent being, namely via machine learning. There, experience
has shown that, rather than let
the machine learn about Lorentz invariance for itself, it is far more
efficient to feed event data to the machine in a Lorentz-invariant
form\footnote{Indeed, as far as we are aware, no computer has yet discovered Lorentz invariance by itself. But, given an arbitrary symmetric metric, a neural network can be trained to converge on the Minkowski metric \cite{NNmetric}.}. There is no reason to expect that permutation invariance should
be any different. Symmetrizing in this way has the related benefit of
preventing the machine chasing wild geese, in the sense of looking for
spurious Lorentz- or permutation-violating signals.\footnote{Of course, this
`benefit' will be considered a disbenefit by readers who are interested
in the possibility that Lorentz invariance is violated, or that, say,
2 protons are not identical; we tactfully suggest that it would be
better for all concerned if they were not to read any further.}
Symmetrizing may even be an astute tactic in situations where the
particles in question are known to be {\em not} identical, but where
one wishes to deliberately blind oneself to the difference between
them, because the associated physics is not under control.  A good
(though politically incorrect) example from the LHC might be a Swiss
proton and a French proton (or rather beams thereof), where one can be
fairly sure that there are observable differences between them, but
one can be equally sure that such differences are not due to
fundamental new physics, but have a rather more mundane, to wit intermural,
origin.

We hope that symmetrizing with respect to permutations in this way
will also be of use in analyses carried out by rather more intelligent
beings. To give just one example, a common method for computing
multi-loop amplitudes in quantum field theory is to first relate them
using integration-by-parts identities \cite{IBP1, IBP2}. These are linear equations
whose coefficients may be written as Lorentz- and
permutation-invariant polynomials in the momenta of external
particles. Thus, in setting up and carrying out such calculations, it would
presumably be useful to know a set of generators of such
polynomials.\footnote{Such permutation invariant polynomials
    may also be of use in analysing correlation functions in
    cosmology, but we will not consider this possibility further here.}

Our goal then in this work will be to generalize Weyl's theorem (namely
supplying an explicit set of generators) to the situation where an
arbitrary subgroup $P \subset S_n$ of the permutation group acts on
the $n$-particles. It would be an insult to Weyl's memory not to do so
in a rigorous fashion, which requires the mathematical machinery of
commutative algebra, the pertinent parts of which we
review in Appendix \ref{sec:Appendix}. But, not least for the benefit of readers who wish to
avoid such unpleasantries, let us first give a more vernacular
statement of the results (such readers may wish to
skip directly  thereafter to the examples giving explicit sets of
generators in \S \ref{sec:Example}.) 

Similar ideas were explored in \cite{Melia}, in the context of classifying higher-dimensional operators in effective scalar field theories. A significant difference there is that one studies the action of permutations on quotient rings with respect to an ideal which features the relation $\sum_i p_i = 0$ (corresponding to an integration-by-parts identity) in addition to the relations $p_i^2 = 0$ studied here (corresponding there to the leading order equations of motion). These additional relations make it difficult to compare our results directly with those in \cite{Melia}, though we hope that some of the results obtained here could nevertheless be usefully applied to the study of that problem. For a rather different approach, see \cite{Thaler}, which studies permutation invariance directly at the level of quantum field theory amplitudes.

\subsection{Non-technical statement of results} \label{sec:NonTech Results}
In layman's terms, Weyl's result is the
statement that every Lorentz invariant polynomial can be obtained by
taking an arbitrary polynomial in variables $y_{ij}$ (where $i,j \in
\{1,\dots,n\}$ and $i\leq j$), and replacing $y_{ij} \mapsto p_i \cdot
p_j$. Our first result is that every
Lorentz and permutation invariant polynomial can be obtained by taking
a permutation invariant polynomial in $y_{ij}$ (where the permutation
group $P$ acts on the indices $i,j$ in the obvious way) and making the same
replacement. 

In a sense, this result {\em is} the
generalization of Weyl's result, but not only is it apparently
completely trivial (though the proof will show it to be not quite so),
but also it is completely useless as it stands, because of the
difficulty of describing the permutation invariant polynomials in
$y_{ij}$. Indeed, while permutations act in the natural way on
the subset $\{y_{ii}\}$ and lead to a simple description of the invariant
polynomials (going back, in the case $P=S_n$, to
Gauss \cite{Gauss}), the action of permutations on $\{y_{i j}|i < j\}$ is
non-standard and a description of the invariants (for the case
$P=S_n$) is unknown for $n\geq 5$
\cite{dk}! Fortunately, such high multiplicities of identical
particles are relatively rare in applications. Our second `result', then, is to describe
and carry out a strategy for finding a set of generators of
the permutation invariant polynomials in $y_{ij}$ for specific cases
of $n$ and $P$, with at most 4 identical particles
(such as for the $pp\to jjj$ example). The strategy uses well-known
methods in
invariant theory, relying crucially on the somewhat arcane
  Cohen-Macaulay property.\footnote{For readers who are not {\em au
    courant}, it is perhaps consoling to note that even
  Macaulay himself professed to being ignorant of this property.}

The list of generators obtained in this way is somewhat lengthy in
practice and so we turn to ways of shortening it. Again, there are
standard ways in invariant theory of doing so, which we describe.\footnote{There is a
  price to be paid for doing so, which we describe shortly.} We also
describe a more {\em ad hoc} method: the observables $p_i \cdot p_i$
for a particle are somewhat redundant, since they return the mass of
the particle (for a jet, we assume that all jet masses are negligible,
since to do otherwise would invalidate the assumption that jets are
identical). As such, we are less interested in invariant polynomials
involving $p_i \cdot p_i$. Unfortunately, one cannot simply throw them
away, because when $n>d$ there are relations between $p_i\cdot p_j$
which mix pairs with $i=j$ and $i\neq j$ (with $n=2$ and $d=1$, for
example, we have that $(p_1 \cdot p_1) (p_2 \cdot p_2) = (p_1 \cdot
p_2)^2$). Our third `result' is to replace this by a kosher procedure
(which is essentially to form a quotient with respect to the ideal generated by
the polynomials $p_i\cdot p_i - m_i^2$, or rather the permutation invariant
combinations thereof) and to provide a set of generators thereof. 

As we will see, these results eventually lead to a manageable set of
generators describing the Lorentz and permutation invariant
polynomials. In the example of $pp \to 3j$, for example, we end up
with a set of 26 generators, given explicitly in Table \ref{table:Gens5}. In fact, this set of generators is minimal in number, so one can do no better. 

\subsection{Technical statement of results} \label{sec:Tech Results}
Let us now give a more technical statement of the results. 
Firstly, it is convenient to regard the momenta as taking values in a
vector space $V \cong \C^{nd}$ over the algebraically-closed field of
complex numbers. Doing so not only leads to simplifications on the
commutative algebra side, but also allows us to replace the Lorentz
group by its complexification $O(d)$. The polynomials in the momenta
then form an algebra,\footnote{In this work, `algebra' will always be understood to
  mean `graded
  algebra over $\C$', unless stated otherwise.} which we denote $\C[V]$ and the
Lorentz-invariant polynomials form a subalgebra $\C[V]^{O(d)}
\subset \C[V]$. A `set of generators' of $\C[V]^{O(d)}$ is equivalent to
a surjective algebra map from some polynomial algebra
to $\C[V]^{O(d)}$. Phrased in these terms, Weyl's theorem is that
there exists such a map $W:\C[y_{ij}] \twoheadrightarrow \C[V]^{O(d)}$, where
$\C[y_{ij}]$ is the polynomial algebra in variables $y_{ij}, i,j \in
\{1,\dots,n\}, i\leq j$, given explicitly on the generators by
$W: y_{ij} \mapsto p_i \cdot p_j$ and extended to an arbitrary polynomial in the obvious way.

Our first result, which follows almost immediately from Weyl's, is
that $W$ restricts to a surjective map between
$\C[y_{ij}]^P \subset \C[y_{ij}]$ and  $\C[V]^{O(d)\times P} \subset
\C[V]^{O(d)\times P}$, the subalgebras that are invariant under
$P\subset S_n$. 
Thus a set of generators of $\C[y_{ij}]^P$ provides us with a set of
generators of the object of interest, $\C[V]^{O(d)\times P}$. Finding
a set of generators of $\C[y_{ij}]^P$ is where the real hard work
begins. Indeed, while the action of $P$ on the subalgebra
$C[y_{ii}]$ is via the natural permutation representation
group, whose invariant algebra is well-understood (a result due to Gauss
in the `worst-case scenario' $P=S_n$ tells us, for example, that
$C[y_{ii}]^{S_n}$ is isomorphic to the polynomial algebra in $n$
variables with degrees $1,\dots,n$), the invariants of the
action of $P$ on the subalgebra $\C[y_{ij} | i<j]$ are rather harder to
describe, with a known description of $\C[y_{ij} | i<j]^{S_n}$ only known for
$n<5$, even though an algorithm is available \cite{dk}.

Thus, we content ourselves with finding generators for
$n$ particle events in which at most 4 particles are identical,  using the fact that the ring of
invariants is Cohen-Macaulay and therefore possesses a Hironaka decomposition. That is, it can be
expressed as a free, finitely-generated module over a polynomial
subalgebra. Thus we may write $\C[y_{ij}]^P = \bigoplus_k \eta_k
\C[\theta_l]$, where $\eta_k$ and $\theta_l$ are polynomials in
$y_{ij}$. Evidently, $\eta_k$ and $\theta_l$ collectively generate
$\C[y_{ij}]^P$ and we refer to them as secondary and primary
generators, respectively. 

There exist algorithms for computing $\eta_k$ and
$\theta_l$, though even modern computers quickly run out of steam (hence
the difficulties when $n \geq 5$).  In this way, we are able to find a
set of generators, whose number is typically rather large (for $pp \to
3j$, for example, we have $10$ primaries and $360$ secondaries for $\C[y_{ij}]^{S_2 \times S_3}$). To pare it
down to a more manageable number, we employ two further
strategies. Firstly, the form of the Hironaka decomposition implies that
the ring multiplication is encoded in the relations $\eta_k \eta_m = \sum_j
f^j_{km} \eta_j, f^n_{km} \in \C[\theta_l]$, and
these can often be used to remove some generators,
which are redundant in the sense that they can be obtained as
algebraic combinations of other generators. (The price to pay is that
the description of the algebra in terms of the remaining generators
becomes more complicated.) Secondly, since
the dot product $p_i\cdot p_i$ does not vary from event to event, being fixed equal to the invariant mass
$m_i^2$, we repeat our construction starting from the quotient ring
$\C[V]/\langle p_i^2-m_i^2 | \forall i \rangle $, showing that there is a
surjection of algebras (which is now no longer graded, since the ideal is
not homogeneous) $\C[y_{ij} | i<j ]^P \twoheadrightarrow
\C[V]^P/(\langle p_k^2-m_k^2  | \forall k \rangle \cap \C[V]^P)$.

We describe the effects of removing parity (which after
  complexification amounts to replacing $O(d)$ by its subgroup $SO(d)$) in \S
  \ref{sec:Parity}. This is conceptually straightforward, in that it
  can be achieved by adding further objects $z_{i_1 \dots i_d}$ to the
  $y_{ij}$, which map under $W$ to contractions of the epsilon tensor
  in $d$ dimensions with $d$ momenta. But in practice, elucidating the structure of
  the corresponding ring of permutation invariants quickly becomes complicated.

Even without permutation invariance, the map to
$\C[V]^{O(d)\times P}$ does not inject for $d >n$ (as the example
given earlier with
$d=1$ and $n=2$
illustrates). This means that there are yet further relations between the
generators of $\C[V]^{O(d)\times P}$ (beyond those in $\C[y_{ij}]^P$), which may be rather obscure\footnote{In the case without
  permutation invariance, the kernel of the map is generated by the
  $d+1$-minors of the matrix $y_{ij}$ (the second fundamental theorem of invariant theory for the orthogonal group).} and which may yet further
frustrate phenomenological analyses. In a follow-up paper, we exploit
the fact that the
algebras $\C[V]^{O(d)\times P}$ are themselves Cohen-Macaulay, meaning
that they too admit a Hironaka decomposition, to describe them
directly and give some explicit examples. Thus each element can
be written uniquely in terms of given primaries $\theta_l$ and secondaries
$\eta_k$  with
a simple multiplication structure  $\eta_k \eta_m =
\sum_j f^j_{km} \eta_j$, which may, for example, be straightforward to
implement on a computer. Unfortunately, we are unable to make much progress
beyond the first non-trivial case, $n=d+1$, but we hope that our
results there may inspire others to try to go further. 
\section{General arguments}
\subsection{Generators for Lorentz and permutation invariants} \label{sec:Gens LPI}
Let a subgroup $P \subset S_n$ of the permutation group act in the
standard way on the indices $i \in \{1, \dots n\}$. This action
induces, in an obvious way, actions on $\{p_i\}$ and $\{ y_{ij} \}$ (with
the obvious rule that we
we replace $y_{ij}$ by $y_{ji}$ if $i>j$) and thence
on $\C[y_{ij}]$, $\C[V]$, and (since the action of permutations
commutes with that of Lorentz transformations) on
$C[V]^{O(d)}$. Moreover, it is easily checked that the Weyl map $W:\C[y_{ij}]
\twoheadrightarrow \C[V]^{O(d)}$ is equivariant with respect to
$P$. That is, given $p \in P$, the diagram 
\begin{equation*}
\begin{tikzcd}
\C[y_{ij}] \arrow[r, twoheadrightarrow, "W"] \arrow[d, "p"] & \C[V]^{O(d)} \arrow[d,"p"] \\ 
\C[y_{ij}] \arrow[r, twoheadrightarrow, "W"] & \C[V]^{O(d)}
\end{tikzcd}
\end{equation*}
commutes. 

From here, we wish to show that $W$ restricts to a surjective map $\C[y_{ij}]^P
\twoheadrightarrow \C[V]^{O(d) \times P}$, so that a set of generators of
$\C[y_{ij}]^P$ furnishes us with a set of generators of $\C[V]^{O(d)
  \times P}$ via evaluating $y_{ij} \mapsto p_i \cdot p_j$.  

To do so, we first note that $W$ sends a $P$-invariant polynomial to a $P$-invariant
polynomial; in other words $W(\C[y_{ij}]^P) \subset \C[V]^{O(d) \times
  P}$ and so there is a well-defined restriction map $W|:\C[y_{ij}]^P \to \C[V]^{O(d) \times
  P}$. It remains to show that the $W|$
map surjects. To do so, let $q \in \C[V]^{O(d) \times
  P} \subset \C[V]^{O(d)}$. Since $W$ is onto, there exists $r \in
\C[y_{ij}]$ such that $W(r) = q$. But $r$ is not necessarily
$P$-invariant, so consider instead $\overline{r} = \frac{1}{p}\sum_{p \in P}
r^p$, where $r^p$ denotes the result of acting on $r$ with $p \in P$. This is $P$-invariant and moreover, we have that
$W|(\overline{r}) = W \left(\frac{1}{p}\sum_{p \in P}
r^p \right) = \frac{1}{p} \sum W(r^p) = \frac{1}{p} \sum (W(r))^p =
\frac{1}{p} \sum q^p = q$, where we used the facts that $W$ is
an algebra map, that $W$ is $P$-equivariant, that $P \subset S_n$ is a
finite group, and that $q$ is $P$-invariant by assumption. Thus $W|$ is onto. 
\subsection{Generators for permutation invariants} \label{sec:Gens PI}
Our next goal is to find a set of generators of the ring
$\C[y_{ij}]^P$, which will in turn provide us with a set of generators
for $\C[V]^{O(d)
  \times P}$. In the case considered by Weyl, where $P$ is the trivial
group, this is a triviality, since $\C[y_{ij}]$ is a polynomial algebra
and so a set of generators (which is moreover a minimal set of
generators) is given by $\{y_{ij}\}$.  In cases where $P$ is not the
trivial group, finding a set of generators is rather harder than it
may first appear. To see why this is the case, consider the `worst
case scenario' $P=S_n$. The group $S_n$ acts reducibly on the subspaces with
bases $\{y_{ii}\}$ and $\{y_{ij}\}$, so there is a well-defined action
(for any $P \subset S_n$, in fact) on the polynomial subalgebras $\C_= := \C[y_{ii}]$ and $\C_<
:=\C[y_{ij} | i< j]$; to begin with, it is helpful to consider
these separately. 

The action of $S_n$ on $\{y_{ii}\}$ is via the natural
permutation representation (in terms of irreducible representations in partition notation it
is $1 \oplus (n-1,1)$) and a complete description of the invariant algebra
$\C_=^{S_n}$ was given by Gauss: it is isomorphic (as a graded
$\C$-algebra) to the polynomial ring in $n$ variables with degrees
$1,\dots, n$. For an explicit isomorphism, one can take {\em e.g.} the symmetric
polynomials $\sum_i y_{ii}, \sum_{i<j} y_{ii}y_{jj},
\dots, \prod_i y_{ii}$ or the power sum polynomials $\sum_i
y^k_{ii}$ with $k \in \{1,\dots n\}$. 

The action of $S_n$ on $\{y_{ij}\}$ is non-standard (in terms of
irreducible representations it is $1 \oplus (n-1,1) \oplus (n-2,2)$ \cite{FultonHarris}). A
description of the invariant algebra is trivial in $n=2,3$, being
given by polynomial algebras in 1 and 3 variables, respectively, but was
only determined relatively recently for $n=4$ \cite{S_4} and is unknown for
$n\geq 5$. It is important to note that the invariant algebra is not a
polynomial algebra for $n\geq 4$. Rather, like any ring of invariants
under the action of a finite group, it has the more general
structure of a Cohen-Macaulay ring \cite{Bruns}. Such rings admit \cite{AlgorithmsinInvTheory} a Hironaka
decomposition as a free, finitely-generated module over a polynomial
subalgebra. Thus we may write $\C[y_{ij}]^P = \bigoplus_k \eta_k
\C[\theta_l]$, where $\eta_k$ and $\theta_l$ are polynomials in
$y_{ij}$. Evidently, $\eta_k$ and $\theta_l$ collectively generate
$\C[y_{ij}]^P$ (of course, we may safely discard the secondary 1) and we refer to them as secondary and primary
generators, respectively. Moreover, we must have that the product of
$\eta_k$ and $\eta_{k^\prime}$ is some linear combination of $\eta_m$s
with coefficients in $\C[\theta_l]$, so we see that the structure of
the algebra is encoded in a simple way. 

Before going further, it is perhaps helpful to give a simple example
of such a ring. One is easily at hand in the form on the ring of
Lorentz and parity invariants of 2 particles in 1 dimension. The
Lorentz and parity transformations reduce to the finite group $\Z/2\Z$
acting on momenta via $p_i \mapsto -p_i$ and so we see that the ring
of invariants admits the Hironaka decomposition $\C[V]^{O(1)} =
1\cdot \C[p_1^2,p_2^2] \oplus p_1p_2 \cdot \C[p_1^2,p_2^2]$, with primaries
$\theta_1 :=p_1^2,
\theta_2 := p_2^2$, secondaries $\eta_1 := 1, \eta_2 := p_1p_2$ and
algebra encoded by $\eta_2^2 = \eta_1 \cdot (\theta_1 \theta_2)$.

Since an explicit description of $\C_<^{S_n}$ is, in general,
unavailable, it is unrealistic to expect one to be available for the
full invariant algebra $\C_\leq^P$ (where we use $\C_\leq$ as a
shorthand to denote the full
$\C[y_{ij}|i \leq j]$). But, since it too has the Cohen-Macaulay property, we
can use the available algorithms to find a Hironaka decomposition in
simple cases. As we will see, the number of primaries and secondaries
that arise in such cases is rather large, so before describing the
algorithms and their outputs explicitly, we first describe a way of
reducing the number of generators, by `removing' the invariant masses
$p_i \cdot p_i$. To do so in a rigorous way requires us to form
quotients of the algebras with respect to the ideal generated by
$p_i^2 - m_i^2$, for all $i$, (or rather its intersection with the invariant ring) which we do in the next Subsection. 
\subsection{Removing invariant masses}
\subsubsection*{Without permutations} \label{sec:Quot-Mass}
Let us warm up by returning to the case considered by Weyl, without
permutation symmetry. Consider the ring formed by taking the quotient
of $\C [V]^{O(d)}$ with respect to the ideal $I$ generated by the
  $O(d)$-invariant elements $p_i^2 -
  m_i^2 $, for all $i$, $\langle p_i^2 -
  m_i^2 | \forall i \rangle$ (where we allow the particle mass-squareds $m_i^2$ to be
  arbitrary complex numbers). We wish to show that there is a
  surjective algebra map\footnote{It is important to note that, unless
  $m_i^2 = 0$ for all $i$, such that $I$ is homogeneous,
  $\C[V]^{O(d)}/I$ is not graded, and so nor is the map.} $\C_< \twoheadrightarrow \C [V]^{O(d)}/I $, such that we can use the $y_{ij}$ with $i<j$ as a set of
  generators. Of course, this result will hardly come as a surprise to
  readers, but making a careful proof in this case will help us to
  avoid potential pitfalls once we add the requirement permutation
  invariance.

The proof has two parts. One part is to show that the
Weyl map $W$ induces a surjective algebra map $\C_\leq/\langle y_{ii} - m_i^2  | \forall i 
\rangle \twoheadrightarrow \C [V]^{O(d)}/ I$. The other part is to
exhibit an algebra isomorphism $\C_\leq/\langle y_{ii} - m_i^2 | \forall i 
\rangle \xrightarrow{\sim} \C_<$.

For the first part, it is enough to note that the map is well-defined
on equivalence classes,
because any element in $\langle y_{ii} - m_i^2 | \forall i 
\rangle $ lands in $I$. (Surjectivity follows automatically from the
surjectivity of $W$.)

For the other part, consider the polynomial algebra $R[x]$ in one variable
over an arbitrary ring $R$. Let $f(x) \in R[x]$, let $r \in R$, and let
$\mathrm{ev}: R[x] \twoheadrightarrow R$ be the evaluation map, {\em
  viz.} the $R$-algebra map defined by $x
\mapsto r$. Since $(x-r)$ is a monic polynomial, by the
division algorithm we have that $f(x) = g(x) \cdot (x-r) + s$, with $g(x)
\in R[x]$ and $s \in R$. Thus $\mathrm{ev} (f(x)) = s$ and $\ker \mathrm{ev} = \langle
x-r \rangle $. By the first isomorphism theorem, $R[x]/\langle x-r
\rangle \xrightarrow{\sim} R$. Now apply this successively to
$\C_{\leq} \cong \C[y_{ij} |(i,j) \neq (1,1)] [y_{11}]$, $\C[y_{ij}
|(i,j) \neq (1,1)] \cong \C[y_{ij} |(i,j) \neq (1,1),(2,2)][y_{22}]$, \&c.
to get the desired result. Equivalently, an explicit isomorphism $\C_\leq/\langle y_{ii} - m_i^2 | \forall i 
\rangle \xrightarrow{\sim} \C_<$ can be obtained from the evaluation map
(which is ungraded, except in the $m_i^2 = 0$ case) from $\C_\leq$
  to $\C_<$ given by
\begin{equation}
\mathrm{ev} : \C_\leq \twoheadrightarrow \C_< :  y_{ii},y_{ij} \mapsto m_i^2,y_{ij}
\end{equation}
whose kernel is indeed $\ker \mathrm{ev} = \langle y_{ii} - m_i^2 | \forall i \rangle$.
\subsubsection*{With permutations} \label{sec:Perm. Quot-Mass}
Now that we have tackled the case without permutations, we turn to address the cases with permutation symmetry. Our goal is to show that there exists a surjective algebra map \footnote{Again, ungraded unless $m_i^2 =0$.} $\C_<^P \twoheadrightarrow \C [V]^{O(d)\times P} / J $ where $J = \langle p_i^2 - m_i^2 | \forall i 
\rangle \cap \C [V]^{O(d) \times P}$. Again, the proof has two parts. One is to show that the restricted Weyl map $W|$ induces a surjective algebra map $\C_\leq^P / J^\prime \twoheadrightarrow \C [V]^{O(d) \times P}/ J$, where $J^\prime = \langle y_{ii} - m_i^2 | \forall i 
\rangle \cap \C_\leq^P$, and the other is to
exhibit an algebra isomorphism $\C_\leq^P / J^\prime \xrightarrow{\sim} \C_<^P$.

For the first part, we begin by showing that the image $W(J^\prime)
\subset J$. For an element $j^\prime \in J^\prime$, $j^\prime  \in \langle y_{ii} - m_i^2 | \forall i  \rangle$ and $j^\prime  \in \C_\leq^P$ by definition. But since the image $W(\langle y_{ii} - m_i^2  | \forall i \rangle) \subset I$, the image $W(j^\prime) \in I$. Furthermore, the element $j^\prime $ is $P$-invariant by assumption and as the map $W$ is $P$-equivariant, the image $W(j^\prime)$ is also $P$-invariant. So, $W(j^\prime) \in J$ and hence $W(J^\prime) \subset J$.\footnote{Actually, $W(J^\prime) = J$, but equality is unnecessary for our purposes.} It is then enough to note that the map is well-defined on the equivalence classes because any element of $J^\prime$ lands in $J$. (Surjectivity again follows from the surjectivity of $W$.)

For the second part, it turns out that the required result follows
from a more general theorem. Suppose that a finite group $G$ acts
reducibly on a vector space $V = X \oplus Z$ and suppose that the
representation carried by $X$ is further reducible, containing the
trivial representation. Let $\{x_i\}$ and $\{z_i\}$, respectively, be bases of
the dual spaces Hom$(X,\C)$ and Hom$(Z,\C)$, respectively, and let $a \in X$ denote a $G$-invariant vector
with components $a_i = x_i(a) \in \C$. 
Further, consider the algebras $R = \mathbb{C}[x_1, \dots, x_m, z_1,
\dots, z_n]$ and $S = \mathbb{C}[z_1, \dots, z_n]$ along with the
evaluation map $\mathrm{ev}: R \twoheadrightarrow S$, $x_i \mapsto a_i$,
with kernel $\langle x_i - a_i  | \forall i \rangle$. We wish to show that there exists an isomorphism of (ungraded if $a_i \neq 0$) algebras $R^G / J \xrightarrow{\sim} S^G$, where $R^G, S^G$ are the $G$-invariant subalgebras of $R,S$ respectively and $J = \langle x_i - a_i | \forall i  \rangle \cap R^G$ is an ideal of $R^G$. 

To do so, we start by explicitly defining the action of $g \in G$
on $h \in R^G$ and $f \in S$ via the reducible representation $\rho: G
\rightarrow GL(V) : g \mapsto \rho_X(g) \oplus \rho_Z(g)$ to be as follows
\begin{align}
& h \mapsto h^g =h(\rho_X(g)x_i, \rho_Z(g) z_i) = h(x_i, z_i) = h, \\
& f \mapsto f^g = f(\rho_Z(g) z_i).
\end{align}
Next, we define the inclusion map $i:R^G \hookrightarrow R$ and compose it with the evaluation map to get the
restricted algebra map $\mathrm{ev}| := \mathrm{ev} \circ i : R^G \rightarrow S$. It can then be checked that the evaluation map $\mathrm{ev}|$ is equivariant with respect to $G$. That is, given $g \in G$, the diagram
\begin{equation*}
\begin{tikzcd}
R^G \arrow[r,"\mathrm{ev}|"] \arrow[d,"g"] & S \arrow[d,"g"] \\ 
R^G \arrow[r,"\mathrm{ev}|"] & S
\end{tikzcd}
\end{equation*}
commutes. Now as the map $\mathrm{ev}|$ is $G$-equivariant, it sends a
$G$-invariant polynomial to a $G$-invariant polynomial; in other words
$\mathrm{ev}|(R^G) \subset S^G$ and so we have a well-defined
restriction map $\mathrm{ev}|: R^G \rightarrow S^G$. It remains to show
that $\mathrm{ev}|$ is surjective. To do so, let $s \in S^G \subset
S$. Since $\mathrm{ev}$ is onto, there exists $r \in R$ such that
$\mathrm{ev}(r) = s$. But $r$ is not necessarily $G$-invariant, so
consider instead $\bar{r} = \frac{1}{|G|}\sum_{g \in G} r^g$, where
again $r^g$ denotes the result of acting on $r$ with $g \in G$. This
is $G$-invariant and we have, furthermore, that
$\mathrm{ev}|(\bar{r}) = \mathrm{ev}| \left( \frac{1}{|G|}\sum_{g \in G}
  r^g \right) = \frac{1}{|G|}\sum_{g \in G} \mathrm{ev}|(r^g) =
\frac{1}{|G|}\sum_{g \in G} \left( \mathrm{ev}|(r) \right)^g = \frac{1}{|G|}\sum_{g \in
  G} s^g = s$, where we have used the fact that $\mathrm{ev}|$ is an
(ungraded for $a_i \neq 0$) algebra map, that $\mathrm{ev}|$ is
$G$-equivariant, that $G$ is a finite group, and that $s$ is
$G$-invariant by assumption. Thus, $\mathrm{ev}|$ is onto. The last
ingredient of the proof is to note that the kernel of the map
$\ker \mathrm{ev}|$ is the restriction of the ideal $\langle x_i - a_i  | \forall i \rangle$ to the $G$-invariant subalgebra $J = \langle x_i - a_i  | \forall i \rangle \cap R^G$. Finally, by the first isomorphism theorem, $R^G / J \xrightarrow{\sim} S^G$. 

In our specific case, the variables $y_{ii}, y_{ij}$ transform
under reducible representations of the permutation group $P$ with the
representation of $y_{ii}$, ($1 \oplus (n-1,1)$), being further reducible
containing the trivial representation. Furthermore, the masses $m_i^2$
clearly form an invariant vector when the particles (and hence the
masses) are identical. Hence, the previous theorem applies and we
have an isomorphism of (ungraded, except in the massless case)
algebras $\C_\leq^P / J^\prime \xrightarrow{\sim} \C_<^P$.
\section{Generators of permutation invariants} \label{sec:gens}
We now describe various results from the theory of invariants which
together may
be used to find sets of generators for the
algebras of permutation invariants, such as $\C_<^P$.  For more
details, see {\em e.g.} \cite{AlgorithmsinInvTheory,dk}. 

Let $K$ be an algebraically-closed field, $V$ a finite-dimensional
vector space over $K$ carrying a representation of a finite group $G$,
and $K[V]$ the polynomial algebra on $V$. The algebra $K[V]$
carries a grading with $(K[V])_0 = K$, which is inherited by the invariant subalgebra
$K[V]^G= \{ f
\in K[V] | \ f^g = f \ \forall g \in G \}$. A result of
Hilbert is that $K[V]^G$ is finitely-generated, while a result of
Noether is that any
finitely-generated graded algebra $R$ with $R_0 = K$ admits a (not
necessarily unique)
homogeneous system of parameters (HSOP). Thus we have that $K[V]^G$ is a
finitely-generated module over $K[\theta_1\dots,\theta_l]$, where the
$\theta_i$ are algebraically-independent. We call the $\theta_i$ {\em primary
  invariants}. In particular, we may write $K[V]^G = \sum_k \eta_k K[\theta_1\dots,\theta_l]$,
where we call the $\eta_j$ {\em secondary invariants}.

Now comes perhaps the most significant result, namely that
$K[V]^G$ is Cohen-Macaulay, which implies that $K[V]^G$ is a {\em
free} (and as we have already seen, finitely-generated) module over
any HSOP. Thus, we in fact have a {\em
  Hironaka decomposition} $K[V]^G = \bigoplus_k \eta_k K[\theta_1\dots,\theta_l]$
and we are able to use the full power of linear algebra. In
particular, each element in $K[V]^G$ can be written uniquely as
$\sum_{j} \eta_j f^j$, where $f^j \in K[\theta_1\dots,\theta_l]$,
and the product of any two secondaries is uniquely given by $\eta_k
\eta_m = \sum_{j} \eta_j f^j_{km}$, where $f^j_{km} \in
K[\theta_1\dots,\theta_l]$. This specifies the multiplication in
$K[V]^G$ unambiguously.

Some simple examples will perhaps be illuminating. When $G$ is the trivial group acting on a
basis vector $x \in \C$, we may set
$\eta_1 =1$ and $\theta_1 =x$, such that $K[V]^G = 1\cdot \C[x]$. But we may
also set $\eta_1=1, \eta_2=x$, and $\theta_1 =x^2$, such that $K[V]^G = 
1\cdot \C[x^2] + x\cdot \C[x^2]$. This already shows that a Hironaka
decomposition is not unique. For a slightly less trivial example,
let  $G$ be the
group $\Z/2\Z$ whose non-trivial element sends basis vectors $x,y \in
\C^2$ to minus themselves. Then we may set  $\eta_1=1, \eta_2=xy$ and
$\theta_1 =x^2, \theta_2 =y^2$.

Clearly, given a Hironaka decomposition, the set containing the
primary and secondary invariants forms a set of generators
of $K[V]^G$, which is what we seek. A Hironaka decomposition can be found by a two-step
  process. The first step is to find an HSOP. It turns out that necessary and sufficient conditions for
a set of homogeneous elements in $K[V]^G$ to form such a system are that they be algebraically independent and that the
locus of points where all elements of strictly positive degree simultaneously vanish is given
by the zero vector in $V$. 

Finding an HSOP has been reduced to an (unwieldy) algorithm \cite{Buchberger, CoxLittleOShea}, but we will
not need it here. Indeed, the group $P$ acts on $\C^P_<$, say (an analogous
result holds for $\C^P_\leq$), by permuting the $y_{ij}$ amongst
themselves; but it is easily shown ({\em cf.} \cite{dk}, Example
2.4.9) that for any permutation subgroup of $S_{n(n-1)/2}$, a HSOP is
given by the $n(n-1)/2$ elementary symmetric polynomials in
$y_{ij}$.

For our purposes, this HSOP is sometimes less than optimal, because it
introduces primary invariants of unnecessarily high degrees, leading
to more secondary invariants (as can
easily be seen by considering the case where $P$ is the trivial group,
such that $\{y_{ij}\}$ is a HSOP, with primary invariants all of
degree 1). An HSOP with primary invariants of lower degrees can be
found by partitioning the $y_{ij}$ into their orbits under $P$
and forming the respective sets of elementary symmetric
polynomials. Again, one may easily show that the union of these forms an HSOP.

Let us make this explicit in our $pp \to 3j$ example. Labelling
  the protons by $4,5$ and the jets by $1,2,3$, we have the following orbits: $\lbrace y_{45} \rbrace, \lbrace y_{12}, y_{13}, y_{23} \rbrace, \lbrace y_{14}, y_{15}, y_{24}, y_{25}, y_{34}, y_{35} \rbrace$. Following our prescription, the HSOP will be 
\begin{alignat}{4} \label{eq:primaries}
& e_{1}(y_{45}),& \qquad e_{1}(y_{12}, y_{13}, y_{23}), & \qquad e_{1}(y_{14}, y_{24}, y_{34}, y_{15}, y_{25}, y_{35}), & \qquad e_{4}(y_{14}, y_{24}, y_{34}, y_{15}, y_{25}, y_{35}),\nonumber \\
& & e_{2}(y_{12}, y_{13}, y_{23}), & \qquad e_{2}(y_{14}, y_{24}, y_{34}, y_{15}, y_{25}, y_{35}), & \qquad e_{5}(y_{14}, y_{24}, y_{34}, y_{15}, y_{25}, y_{35}), \nonumber \\
& & e_{3}(y_{12}, y_{13}, y_{23}), & \qquad e_{3}(y_{14}, y_{24}, y_{34}, y_{15}, y_{25}, y_{35}),  &\qquad e_{6}(y_{14}, y_{24}, y_{34}, y_{15}, y_{25}, y_{35}).
\end{alignat}
Having found an HSOP, we turn to the second step in finding a
  Hironaka decomposition, which is to find the corresponding secondary
  invariants. A first observation is that one can read off the degrees of
  the secondary invariants by comparing the Hilbert series computed using Molien's formula
\begin{equation}
H(K[V]^G, t) = \frac{1}{|G|} \sum_{g \in G} \frac{1}{\mathrm{det}(1 - t \cdot \rho_g)}
\end{equation}
(where $\rho_g$ is the linear operator representing $g \in G$) to the
form corresponding to the Hironaka decomposition, {\em viz.}
\begin{equation}
H \left( \bigoplus \eta K[\theta], t \right) = \frac{1 + \sum_k S_k t^k}{\prod_l (1-t^l)^{P_l}}.
\end{equation}
where there are $S_k$ secondaries at degree $k$ and $P_l$
primaries at degree $l$. 
By way of illustration, Table \ref{table:Hilbert Series} lists the Hilbert series for a few
  of the algebras that we are interested in.
\bgroup
\def\arraystretch{2}%
\begin{table}
\begin{center}
\begin{tabular}{ |c|c| } 
\hline
& $\C_<^P$  \\ 
\hline
$n = 4, \text{ with } S_1$  & $\frac{1}{(1-t)^{6}}$  \\
\hline
$n = 4, \text{ with } S_2 \times S_2$  & $\frac{1+t^3}{(1-t)^3 \left(1-t^2\right)^3}$ \\ 
\hline
$n = 5, \text{ with } S_1$  & $\frac{1}{(1-t)^{10}}$  \\ 
\hline
$n = 5, \text{ with } S_2 \times S_3$  & $\frac{1+t^2+6 t^3+8 t^4+6 t^5+12 t^6+14 t^7+9 t^8+8 t^9+5 t^{10}+2 t^{11}}{(1-t)^3
   \left(1-t^2\right)^4 \left(1-t^3\right)^2 \left(1-t^6\right)}$ \\ 
\hline
\end{tabular}
\caption{\label{table:Hilbert Series} The Hilbert series of some
  relevant invariant algebras.}
\end{center}
\end{table}
\egroup 

The secondaries may now be found via the following algorithm
  \cite{dk}, employing
  a Groebner basis\footnote{Readers unfamiliar with these may wish to consult \cite{CoxLittleOShea} for a gentle introduction.} $\mathcal{G}$ for the ideal $\langle
  \theta_1,\dots,\theta_l \rangle  \subset K[V]^G$ generated by the primary invariants:
\begin{itemize}
\item Read off the degrees of secondaries $d_1, \dots d_m$ from the Hilbert series.
\item For $i = 1, \dots ,m$ perform the following two steps:
\subitem - Calculate a basis of the homogeneous component $K[V]_{d_i}^G$ (invariant polynomials of degree $d_i$).
\subitem - Select an element $\eta_i$ from this basis such that the
normal form $\text{NF}_{\mathcal{G}}(\eta_i)$ (remainder on division
by the Groebner basis) is non-zero and is not in the $K$-vector space generated by the polynomials $\text{NF}_{\mathcal{G}}(\eta_1), \dots ,\text{NF}_{\mathcal{G}}(\eta_{i-1})$.
\item The invariants $\eta_1, \dots ,\eta_k$ are the required secondary invariants.
\end{itemize}
A version of this algorithm is implemented in \texttt{Macaulay2} \cite{M2} (and other computer packages).

\section{Redundancies} \label{sec:Redundancies}

In the previous Section, we described a systematic construction of a
Hironaka decomposition, and {\em ergo} a set of generators, for
$\C_<^P$ (an analogous construction applies for $\C_\leq^P$). Unfortunately, the number of generators is rather large in all but
the simplest cases. For the purpose of carrying out phenomenological
analyses, one would like to have a set of generators that is as
minimal as possible, in the sense of reducing both the number of
generators and their degrees. In this Section, we will see that
  such a reduction is indeed possible, and leads to a set of
  generators whose cardinality is minimal (the degrees of the
  generators in such a set is moreover fixed). Unfortunately, the
  number of generators in such a set is still rather large. But this
  is the best one can do.

The reduction may be achieved (at the cost of destroying the neat encoding
of the algebraic
structure in the
  Hironaka decomposition, which may in itself be useful for phenomenological
analyses) via the following
algorithm: For a set of generators $S$, choose an element $f \in S$
and set up a general element of the same degree as $f$ in the algebra
generated by $S \setminus {f}$ with unknown coefficients. Equate it to
$f$ and extract the corresponding system of linear equations by
comparison of coefficients. The system is solvable if and only if $f$
can be omitted from $S$. It turns out \cite{dk}, though we will
  not show it here, that this procedure
leads to a set of algebra generators whose cardinality is minimal; the
degrees of the resulting generators are, moreover, uniquely
determined.

It seems that we are home and dry, but there is one remaining issue:
although the problem of finding the secondary generators is solved
algorithmically, in most non-trivial cases, it is highly
inefficient. Even modern computers using state-of-the-art
algorithms start struggling with Hironaka decompositions containing
more than a few hundred secondaries. Our only hope is if we can
somehow get away with finding some, but not all, of  the secondaries
before using the elimination procedure just described. This hope can
be realised by use of arguments going back to Noether, who showed
that the maximal degree of an algebra generator in a minimal set is
$\leq |G|$. When $G$ is
non-cyclic (so $P \neq S_1,S_2$ in the case at hand), Noether's bound
can be improved to $\frac{3}{4} |G|$ if $|G|$ is even and
$\frac{5}{8} |G|$ if $|G|$ is odd \cite{Domokos}.\footnote{In our $pp \to
  3j$ example, we have $3|G|/4 = 9$, which comfortably exceeds our
  highest degree primary, of degree 6; we will see in the next Section that in fact the
  highest degree in a minimal set of generators is in fact 6.} Therefore, we only need to find the secondaries up to
these bounds before discarding the redundant generators using the
process outlined above. Of course, in many cases these bounds are practically useless; the order of $S_n$ is $n!$. But for physically relevant examples such as $S_2 \times S_3$, they reduce the computation time significantly.

\section{Examples} \label{sec:Example}

In this Section, we will apply the aforementioned techniques to find sets of generators for common examples of phenomenological interest. 

\subsection{$pp \to jj$}
A common scattering problem is the two protons to two jets, $pp
\to jj$, though of course $jj$ could be any two objects that we do not want to or cannot distinguish, which corresponds to the $n = 4$ with $S_2 \times S_2$
case. 

First, we find the primaries using our prescription. The invariant subspaces are $\{ y_{12} \}, \{ y_{34} \}, \\ \{ y_{13},y_{14}, y_{23}, y_{24} \}$, and therefore we take the primaries to be 
\begin{alignat*}{3}
& e_{1}(y_{12}), & \qquad e_{1}(y_{13},y_{14}, y_{23}, y_{24}), & \qquad e_{3}(y_{13},y_{14}, y_{23}, y_{24}),  \\
& e_{1}(y_{34}), & \qquad e_{2}(y_{13},y_{14}, y_{23}, y_{24}), & \qquad e_{4}(y_{13},y_{14}, y_{23}, y_{24}).
\end{alignat*}
We can already see directly from the improved Noether bound
  (which is $\frac{3}{4}(2!)(2!) = 3$ in this case) that these generators cannot be part of a
  minimal set. To read off the degrees of secondaries, we write the
  Hilbert series in Table \ref{table:Hilbert Series} in the form
\begin{align*}
H(\C[y_{ij}]^{S_2 \times S_2}, t) = \frac{1+2t^2 +2t^4 + t^6}{(1-t)^3 (1-t^2) (1-t^3)(1-t^4)}.
\end{align*}
Next, we use the algorithm to compute the secondaries. Using the
bound, we only need to find the secondaries up to degree 3. Once
found, we can start eliminating redundancies from the union of
primaries and secondaries in the fashion described in Section \S
\ref{sec:Redundancies}. Once this is done, we are left with a set of
$7$ minimal algebra generators given in Table \ref{table:Gens4}.
\begin{longtable}{| p{0.7\textwidth} |} 
\hline
Degree = 1 \\
\hline
$ g_{11} = y_{12}, $\\
$ g_{12} = y_{34}, $\\
$ g_{13} = y_{13}+y_{14}+y_{23}+y_{24}, $\\
\hline
Degree = 2 \\
\hline
$ g_{21} = y_{13} y_{23}+y_{14} y_{24}, $\\
$ g_{22} = y_{13} y_{14}+y_{23} y_{24}, $\\
$ g_{23} = y_{13} y_{14}+y_{13} y_{23}+y_{14} y_{23}+y_{13} y_{24}+y_{14} y_{24}+y_{23} y_{24}, $\\
\hline
Degree = 3 \\
\hline
$ g_{31} = y_{13} y_{14} y_{23}+y_{13} y_{14} y_{24}+y_{13} y_{23} y_{24}+y_{14} y_{23} y_{24}$ \\
\hline
\caption{\label{table:Gens4} Table of generators for $n=4$ with $S_2 \times S_2$.}
\end{longtable}

\subsection{$pp \to jjj$}
We now ramp up the level of complexity, by considering $pp \to
  jjj$, which corresponds to the $n = 5$ with $P =S_2 \times S_3$ case.

The set of primaries was already given in equation \ref{eq:primaries} of Section \S
\ref{sec:gens}. Comparing to
the Hilbert series in Table \ref{table:Hilbert Series}, we see that they are again non-optimal and we need to write the Hilbert series in the modified form
\begin{align*}
& H(\C[y_{ij}]^{S_2 \times S_3}, t) = \\ 
& \frac{1+3 t^2+6 t^3+12 t^4+17 t^5+32 t^6+35 t^7+47 t^8+48 t^9+49 t^{10}+38 t^{11}+34
   t^{12}+19 t^{13}+12 t^{14}+5 t^{15}+2 t^{16}}{(1-t)^3 \left(1-t^2\right)^2
   \left(1-t^3\right)^2 \left(1-t^4\right) \left(1-t^5\right) \left(1-t^6\right)}.
\end{align*}
Using the algorithm to find
the secondaries up to degree $\frac{3}{4}(2!)(3!) = 9$ and eliminating redundancies, we are left with a set of $26$ minimal algebra generators. Table \ref{table:Gens5} contains the explicit list.

\begin{longtable}{| p{\textwidth} |} 
\hline
Degree = 1 \\
\hline
$ g_{11} = \mybox{y_{12}, } $ \\
$ g_{12} = \mybox{y_{34}+y_{35}+y_{45}, } $ \\
$ g_{13} = \mybox{y_{13}+y_{14}+y_{15}+y_{23}+y_{24}+y_{25}, } $ \\ \\
\hline
Degree = 2 \\
\hline
$ g_{21} = \mybox{y_{13} y_{23}+y_{14} y_{24}+y_{15} y_{25} , } $ \\ 
$ g_{22} = \mybox{y_{34} y_{35}+y_{34} y_{45}+y_{35} y_{45}, } $ \\ 
$ g_{23} = \mybox{y_{13} y_{14}+y_{13} y_{15}+y_{14} y_{15}+y_{23} y_{24}+y_{23} y_{25}+y_{24} y_{25} , } $ \\
$ g_{24} = \mybox{y_{13} y_{34}+y_{14} y_{34}+y_{23} y_{34}+y_{24} y_{34}+y_{13} y_{35}+y_{15} y_{35}+y_{23} y_{35}+y_{25} y_{35}+y_{14} y_{45}+y_{15} y_{45}+y_{24} y_{45}+y_{25} y_{45} , } $ \\
$ g_{25} = \mybox{y_{13} y_{14}+y_{13} y_{15}+y_{14} y_{15}+y_{13} y_{23}+y_{14} y_{23}+y_{15} y_{23}+y_{13} y_{24}+y_{14} y_{24}+y_{15} y_{24}+y_{23} y_{24}+y_{13} y_{25}+y_{14} y_{25}+y_{15} y_{25}+y_{23} y_{25}+y_{24} y_{25}, } $ \\ \\
\hline
Degree = 3 \\
\hline
$ g_{31} = \mybox{y_{34} y_{35} y_{45} , } $ \\
$ g_{32} = \mybox{y_{13} y_{23} y_{34}+y_{14} y_{24} y_{34}+y_{13} y_{23} y_{35}+y_{15} y_{25} y_{35}+y_{14} y_{24} y_{45}+y_{15} y_{25} y_{45} , } $ \\
$ g_{33} = \mybox{y_{13} y_{14} y_{34}+y_{23} y_{24} y_{34}+y_{13} y_{15} y_{35}+y_{23} y_{25} y_{35}+y_{14} y_{15} y_{45}+y_{24} y_{25} y_{45} , } $ \\
$ g_{34} = \mybox{y_{13} y_{34}^2+y_{14} y_{34}^2+y_{23} y_{34}^2+y_{24} y_{34}^2+y_{13} y_{35}^2+y_{15} y_{35}^2+y_{23} y_{35}^2+y_{25} y_{35}^2+y_{14} y_{45}^2+y_{15} y_{45}^2+y_{24} y_{45}^2+y_{25} y_{45}^2 , } $ \\ 
$ g_{35} = \mybox{y_{13}^2 y_{34}+y_{14}^2 y_{34}+y_{23}^2 y_{34}+y_{24}^2 y_{34}+y_{13}^2 y_{35}+y_{15}^2 y_{35}+y_{23}^2 y_{35}+y_{25}^2 y_{35}+y_{14}^2 y_{45}+y_{15}^2 y_{45}+y_{24}^2 y_{45}+y_{25}^2 y_{45} , } $ \\
$ g_{36} = \mybox{y_{13}^2 y_{23}+y_{13} y_{23}^2+y_{14}^2 y_{24}+y_{14} y_{24}^2+y_{15}^2 y_{25}+y_{15} y_{25}^2 , } $ \\
$ g_{37} = \mybox{y_{13}^2 y_{14}+y_{13} y_{14}^2+y_{13}^2 y_{15}+y_{14}^2 y_{15}+y_{13} y_{15}^2+y_{14} y_{15}^2+y_{23}^2 y_{24}+y_{23} y_{24}^2+y_{23}^2 y_{25}+y_{24}^2 y_{25}+y_{23} y_{25}^2+y_{24} y_{25}^2 , } $ \\
$ g_{38} = \mybox{y_{13} y_{14} y_{15}+y_{13} y_{14} y_{23}+y_{13} y_{15} y_{23}+y_{14} y_{15} y_{23}+y_{13} y_{14} y_{24}+y_{13} y_{15} y_{24}+y_{14} y_{15} y_{24}+y_{13} y_{23} y_{24}+y_{14} y_{23} y_{24}+y_{15} y_{23} y_{24}+y_{13} y_{14} y_{25}+y_{13} y_{15} y_{25}+y_{14} y_{15} y_{25}+y_{13} y_{23} y_{25}+y_{14} y_{23} y_{25}+y_{15} y_{23} y_{25}+y_{13} y_{24} y_{25}+y_{14} y_{24} y_{25}+y_{15} y_{24} y_{25}+y_{23} y_{24} y_{25} , } $ \\ \\
\hline
Degree = 4 \\
\hline
$ g_{41} = \mybox{y_{13}^2 y_{23}^2+y_{14}^2 y_{24}^2+y_{15}^2 y_{25}^2 , } $ \\
$ g_{42} = \mybox{y_{13} y_{23} y_{34}^2+y_{14} y_{24} y_{34}^2+y_{13} y_{23} y_{35}^2+y_{15} y_{25} y_{35}^2+y_{14} y_{24} y_{45}^2+y_{15} y_{25} y_{45}^2 , } $ \\
$ g_{43} = \mybox{y_{13} y_{14} y_{34}^2+y_{23} y_{24} y_{34}^2+y_{13} y_{15} y_{35}^2+y_{23} y_{25} y_{35}^2+y_{14} y_{15} y_{45}^2+y_{24} y_{25} y_{45}^2 , } $ \\
$ g_{44} = \mybox{y_{13}^2 y_{23} y_{34}+y_{13} y_{23}^2 y_{34}+y_{14}^2 y_{24} y_{34}+y_{14} y_{24}^2 y_{34}+y_{13}^2 y_{23} y_{35}+y_{13} y_{23}^2 y_{35}+y_{15}^2 y_{25} y_{35}+y_{15} y_{25}^2 y_{35}+y_{14}^2 y_{24} y_{45}+y_{14} y_{24}^2 y_{45}+y_{15}^2 y_{25} y_{45}+y_{15} y_{25}^2 y_{45} , } $ \\
$ g_{45} = \mybox{y_{13}^2 y_{15} y_{34}+y_{14}^2 y_{15} y_{34}+y_{23}^2 y_{25} y_{34}+y_{24}^2 y_{25} y_{34}+y_{13}^2 y_{14} y_{35}+y_{14} y_{15}^2 y_{35}+y_{23}^2 y_{24} y_{35}+y_{24} y_{25}^2 y_{35}+y_{13} y_{14}^2 y_{45}+y_{13} y_{15}^2 y_{45}+y_{23} y_{24}^2 y_{45}+y_{23} y_{25}^2 y_{45} , } $ \\
$ g_{46} = \mybox{y_{13}^2 y_{14} y_{23}+y_{13}^2 y_{15} y_{23}+y_{13} y_{14}^2 y_{24}+y_{14}^2 y_{15} y_{24}+y_{13} y_{23}^2 y_{24}+y_{14} y_{23} y_{24}^2+y_{13} y_{15}^2 y_{25}+y_{14} y_{15}^2 y_{25}+y_{13} y_{23}^2 y_{25}+y_{14} y_{24}^2 y_{25}+y_{15} y_{23} y_{25}^2+y_{15} y_{24} y_{25}^2 , } $ \\
$ g_{47} = \mybox{y_{13} y_{14} y_{15} y_{23}+y_{13} y_{14} y_{15} y_{24}+y_{13} y_{14} y_{23} y_{24}+y_{13} y_{15} y_{23} y_{24}+y_{14} y_{15} y_{23} y_{24}+y_{13} y_{14} y_{15} y_{25}+y_{13} y_{14} y_{23} y_{25}+y_{13} y_{15} y_{23} y_{25}+y_{14} y_{15} y_{23} y_{25}+y_{13} y_{14} y_{24} y_{25}+y_{13} y_{15} y_{24} y_{25}+y_{14} y_{15} y_{24} y_{25}+y_{13} y_{23} y_{24} y_{25}+y_{14} y_{23} y_{24} y_{25}+y_{15} y_{23} y_{24} y_{25} , } $ \\ \\
\hline
Degree = 5 \\
\hline
$ g_{51} = \mybox{y_{13}^2 y_{23}^2 y_{34}+y_{14}^2 y_{24}^2 y_{34}+y_{13}^2 y_{23}^2 y_{35}+y_{15}^2 y_{25}^2 y_{35}+y_{14}^2 y_{24}^2 y_{45}+y_{15}^2 y_{25}^2 y_{45} } $ \\
$ g_{52} = \mybox{y_{13} y_{14} y_{15} y_{23} y_{24}+y_{13} y_{14} y_{15} y_{23} y_{25}+y_{13} y_{14} y_{15} y_{24} y_{25}+y_{13} y_{14} y_{23} y_{24} y_{25}+y_{13} y_{15} y_{23} y_{24} y_{25}+y_{14} y_{15} y_{23} y_{24} y_{25} , } $ \\ \\
\hline
Degree = 6 \\
\hline
$ g_{61} = \mybox{y_{13} y_{14} y_{15} y_{23} y_{24} y_{25} , } $ \\ \\
\hline
\caption{\label{table:Gens5} Table of generators for $n=5$ with $S_2 \times S_3$.}
\end{longtable}

\section{Parity} \label{sec:Parity}
Finally, we briefly discuss the more general case where parity is not
a symmetry. Weyl showed that a generating set of Lorentz invariants in $d$
dimensions is given by the dot products, along with all the
possible contractions of momenta with the anti-symmetric $d$
dimensional Levi-Civita epsilon tensor\footnote{There are, of course,
  relations between the Levi-Civita tensors and the dot products,
  namely the product of two epsilon tensors contracted with some
  momenta $p_i$ is equal to the corresponding minor of the $p_i \cdot
  p_j$ matrix.}. To include these extra generators in our discussion,
one could add some extra variables $z_{i_1, \cdots, i_d}$ which
transform in a similar (anti-symmetric) manner to the epsilons under
the action of the permutation group and are mapped to the epsilons in
the appropriate way under the Weyl map. One then needs to study
the algebra $\C[y_{ij}, z_{i_1, \cdots, i_d}]^P$ and find its Hironaka
decomposition and consequently a set of minimal algebra
generators. The first challenge one runs into in trying to do so
is the difficulty in finding a suitable HSOP. Since the
elements in $P$ act on $z_{i_1, \cdots, i_d}$ by permutation, an HSOP
is given by the elementary symmetric polynomials in $z_{i_1, \cdots,
  i_d}$, but the degrees of the resulting generators are prohibitively large, with a consequent slew of secondaries. Given the inefficiencies
of current algorithms, which already struggle with the simpler case of
$\C_<^P$, it seems unlikely that one will be able to find a minimal
set of generators in this way, in all but the simplest cases. Most likely,
a more sophisticated approach that takes into account the relations
between dot products and epsilons is needed. We leave this for future work.

\section{Discussion} 

In this work, we have developed a systematic method which produces sets of minimal algebra generators for the Lorentz and permutation invariant polynomials using tools of invariant theory, generalising results obtained by Weyl in the absence of permutation invariance. Our method results in manageable sets of generators for phenomenologically-relevant examples, at least when the number of particles is sufficiently small, and we hope that the results will prove to be useful in future phenomenological analyses. 

Our work has several shortcomings. One is that it does not address redundancies that occur in sufficiently low spacetime dimensions. Another is that we have failed to make substantial progress in the case where parity is not a symmetry. A third problem is that our generators are not able to fully separate the orbits\footnote{To give a somewhat trivial example, the invariant $p \cdot p$ is unable to separate the orbits with $p \cdot p =0 $ and with either $p=0$ or $p \neq 0$.}, which is certainly a useful thing to do from a physicist's point of view (for example in searching for parity violating LHC signals, as explored in \cite{Lester}). We hope to address all of these deficits in future work \cite{LPIPII,LPIPIII}.

\appendix
\setcounter{secnumdepth}{0}
\section{Appendix} \label{sec:Appendix}
Here we recall some relevant definitions (of terms in italics) and results from commutative
algebra (see, {\em e.g.} \cite{atiyahmacdonald,altmankleinman}, for
more details). The most important concepts are those of a {\em ring} and an
{\em algebra}, and the corresponding structure-preserving {\em maps} between them.

A {\em ring} $R$ (which for our purposes will always be a commutative ring with unit)
is an Abelian group (with addition $+$, identity $0$, and element $r
\in R$ having inverse $-r$) that is also a commutative monoid (with
multiplication $\cdot$, which we often omit, and identity $1$), such
that $\cdot$ is distributive over $+$. An example is the ring $\Z$ of
integers.

A {\em ring map} $f:R \to S$ (which we sometimes write less explicitly as $R \to S$) is a map that preserves sums, products, and
$1$. A {\em ring isomorphism} $R \xrightarrow{\sim} S$ is a bijective ring map.

An {\em$R$-algebra} (or {\em algebra} for short) is a ring $S$ equipped with a ring map $f:R \to S$. An
example is the polynomials in one variable $R[x]$ over a ring $R$
(where the ring map is $r \mapsto rx^0$). Given $R$-algebras $S$ and
$T$ with structure maps $f,g$ (respectively), an {\em$R$-algebra map}
is a ring map $h:S
\to T$ such that $h \circ f = g$. 

Given an $R$-algebra $S$, the subalgebra $R[\{s_\lambda|\lambda \in \Lambda\}]$ {\em generated} by $s_\lambda \in
S$ is the smallest $R$-subalgebra that contains them. It consists of all
polynomial combinations of the $s_\lambda$ with coefficients in
$R$. If there exist $s_1,\dots,s_n \in S$ such that $S = R[s_1,\dots,s_n]$, we say
that $S$ is {\em finitely-generated} (as an $R$-algebra). 

The {\em kernel} $\ker f$ of a ring map $f$  is $f^{-1}(0)$. An {\em ideal} $I \subset R$
is the kernel of a ring map. Equivalently, an ideal contains $0$ and is such that given
$a,b \in I$ and $r \in R$, $a+b \in I$ and $ar \in I$ (indeed, this is the
kernel of the map $R \to R/I$ that sends $r$ to the equivalence class
$r+I$, the set of which forms the {\em quotient ring} $R/I$). The {\em first isomorphism
theorem} states that $R/\ker f \xrightarrow{\sim} \im f$. 

The ideal $\langle r_\lambda | \forall \lambda \in \Lambda \rangle$ generated by $r_\lambda \in R$ for some set $\Lambda$, is the smallest ideal in $R$ that contains the
$r_\lambda$. A {\em
  field} is a ring in which $\langle 0 \rangle $ is a {\em maximal
  ideal}, that is, is not contained in any proper ideal. Equivalently,
$1 \neq 0$ and every non-zero element is a {\em unit}, that is has a
multiplicative inverse.

An {\em $R$-module} (or just {\em module}) $M$ is an Abelian group (written additively)
together with a {\em scalar multiplication} $R \times M \to M: (r,m) \mapsto rm$ that is
distributive over the addition in both $R$ and $M$, is associative,
and is such that $1m \mapsto m$. An ideal in $R$ and an $R$-algebra
are both examples of $R$-modules. 

We say that a subset
$\{m_{\lambda} |\lambda \in \Lambda\} \subset M$ {\em generates} $M$ (as a
module) if $M$ is the smallest submodule of $M$ that contains
$\{m_{\lambda} \}$. We say that $M$ is {\em finitely-generated} if there exists
a finite set of generators. We say that  the $m_{\lambda}$ are {\em free} if
$\sum_\lambda r_\lambda m_\lambda = 0 \implies r_\lambda = 0$, for all
$\lambda$ and that they are a {\em basis} if they also generate $M$. A {\em free
module} is one that has a basis.

A ring $R$ is {\em graded} if we can write it as a direct sum $R
= \bigoplus_{n \in \mathbb{N}} R_n$ of subgroups $R_n$ (in
fact $R_0$ is always a subring) such that $R_n R_m \subset
R_{n+m}$. A {\em homogeneous element (of degree $n$)} is an element
belonging to some factor (or specifically to the factor $R_n$). An algebra is graded if it is graded as a ring. 

Given a graded algebra $R$ over a field $K$ with $R_0=K$, a {\em
  homogeneous system of parameters} is a set of homogeneous elements
$\theta_1,\dots, \theta_m \in R$ which are algebraically independent
and are such that $R$ is a finitely-generated module over $K[\theta_1,\dots, \theta_m]$.

For a finitely-generated graded $K$-algebra $R =
\bigoplus_{i=0}^\infty R_i$, we define the Hilbert series $H(R, t)$ as the formal power series
$$
H(R, t) = \sum_{i = 0}^\infty \mathrm{dim}(R_i) t^i
$$
where $\mathrm{dim}(R_i)$ is the dimension of the (homogeneous)
vector space $R_i$.

\section*{Acknowledgements}
We thank Scott Melville and other members of the Cambridge Pheno Working Group for
helpful advice and comments. 
This work has been partially supported by STFC consolidated grants
ST/P000681/1 and 
ST/S505316/1. WH is supported by the Cambridge Trust. BG thanks King's
College, Cambridge, and the University of Canterbury, New Zealand,
where part of this work was carried out.

\bibliographystyle{JHEP-2}
\bibliography{generators}

\providecommand{\href}[2]{#2}\begingroup\raggedright\begin{thebibliography}{10}

\bibitem{Weyl}
H.~Weyl, {\em {The classical groups: their invariants and representations}}.
\newblock Princeton University Press, 1966.

\bibitem{NNmetric}
A.~Butter, G.~Kasieczka, T.~Plehn and M.~Russell, {\it Deep-learned top tagging
  with a lorentz layer},  {\em {SciPost} Physics} (2018).

\bibitem{IBP1}
F.~Tkachov, {\it A theorem on analytical calculability of 4-loop
  renormalization group functions},  {\em Physics Letters B} (1981).

\bibitem{IBP2}
A.~Kotikov, {\it Differential equations method. new technique for massive
  {Feynman} diagram calculation},  {\em Physics Letters B} (1991).

\bibitem{Melia}
B.~Henning, X.~Lu, T.~Melia and H.~Murayama, {\it Operator bases,
  {$S$}-matrices, and their partition functions},  {\em Journal of High Energy
  Physics} (2017).

\bibitem{Thaler}
P.~T. Komiske, E.~M. Metodiev and J.~Thaler, {\it Cutting multiparticle
  correlators down to size},  {\em Physical Review D} (2020).

\bibitem{Gauss}
C.~F. Gauss, {\em {Carl Friedrich Gauss Werke V3 (1876) (German Edition)}}.
\newblock Kessinger Publishing, LLC, 2010.

\bibitem{dk}
H.~Derksen and G.~Kemper, {\em {Computational invariant theory}}.
\newblock Springer Berlin Heidelberg, 2002.

\bibitem{FultonHarris}
W.~Fulton, W.~Harris and J.~Harris, {\em {Representation theory: a first
  course}}.
\newblock Springer New York, 1991.

\bibitem{S_4}
H.~Aslaksen, S.~P. Chan and T.~Gulliksen, {\it Invariants of {$S_4$} and the
  shape of sets of vectors},  {\em Applicable Algebra in Engineering,
  Communication and Computing} (1996).

\bibitem{Bruns}
W.~Bruns and H.~J. Herzog, {\em {Cohen-Macaulay rings}}.
\newblock Cambridge University Press, 1998.

\bibitem{AlgorithmsinInvTheory}
P.~Paule and B.~Sturmfels, {\em {Algorithms in invariant theory}}.
\newblock Springer Vienna, 2008.

\bibitem{Buchberger}
B.~Buchberger, {\it A theoretical basis for the reduction of polynomials to
  canonical forms},  {\em {ACM} {SIGSAM} Bulletin} (1976).

\bibitem{CoxLittleOShea}
D.~Cox, J.~Little and D.~O'Shea, {\em {Ideals, varieties, and algorithms: an
  introduction to computational algebraic geometry and commutative algebra}}.
\newblock Springer New York, 2008.

\bibitem{M2}
D.~R. Grayson and M.~E. Stillman, ``Macaulay2, a software system for research
  in algebraic geometry.'' Available at
  \url{http://www.math.uiuc.edu/Macaulay2/}.

\bibitem{Domokos}
M.~Domokos and P.~Hegedus, {\it Noether's bound for polynomial invariants of
  finite groups},  {\em Archiv der Mathematik} (2000).

\bibitem{Lester}
C.~G. Lester and M.~Schott, {\it Testing non-standard sources of parity
  violation in jets at the lhc, trialled with cms open data},  {\em Journal of
  High Energy Physics} (2019).

\bibitem{LPIPII}
B.~Gripaios, W.~Haddadin and C.~Lester, {\it Lorentz and permutation invariants
  of particles {II}}, . To appear.

\bibitem{LPIPIII}
B.~Gripaios, W.~Haddadin and C.~Lester, {\it Lorentz and permutation invariants
  of particles {III}}, . To appear.

\bibitem{atiyahmacdonald}
M.~Atiyah and I.~MacDonald, {\em {Introduction to commutative algebra}}.
\newblock Avalon Publishing, 1994.

\bibitem{altmankleinman}
S.~Kleiman and A.~Altman, {\em {A term of commutative algebra}}.
\newblock Worldwide Center of Mathematics, LLC, 2013.

\end{thebibliography}\endgroup


\providecommand{\href}[2]{#2}\begingroup\raggedright\endgroup

\end{document}